\documentclass[aps,prc,twocolumn,superscriptaddress,showpacs,nofootinbib]{revtex4-1}
\usepackage{graphicx,amsmath,amssymb,bm,color,physics,rotating}
\usepackage{epstopdf}

\usepackage{hyperref}
\usepackage{url}

\newcommand{\beq}{\begin{equation}}
\newcommand{\eeq}{\end{equation}}

\newcommand{\RomanNumeralCaps}[1]
    {\MakeUppercase{\romannumeral #1}}

\begin{document}

\title{Clustering of Four-Component Unitary Fermions}

\author{William G. Dawkins}
\affiliation{Department of Physics, University of Guelph, Guelph, Ontario N1G 2W1, Canada}
\author{J. Carlson}
\affiliation{Theoretical Division, Los Alamos National Laboratory, Los Alamos, New Mexico 87545, USA}
\author{U. van Kolck}
\affiliation{Institut de Physique Nucl\'eaire, CNRS-IN2P3, Universit\'e Paris-Sud, Universit\'e Paris-Saclay, 91406 Orsay, France}
\affiliation{Department of Physics, University of Arizona, Tucson, Arizona 85721, USA}
\author{Alexandros Gezerlis}
\affiliation{Department of Physics, University of Guelph, Guelph, Ontario N1G 2W1, Canada}

\begin{abstract}
\textit{Ab initio} nuclear physics tackles the problem of strongly interacting four-component fermions. The same setting could foreseeably be probed experimentally in ultracold atomic systems, where two- and three-component experiments have led to major breakthroughs in recent years. Both due to the problem's 
inherent interest and as a pathway to nuclear physics, in this Letter we study four-component fermions at unitarity via the use of quantum Monte Carlo methods. We explore novel forms of the trial wave function and find one which leads to a ground state of the eight-particle system whose energy is almost equal to that of 
two four-particle systems.
We investigate the clustering properties involved and also extrapolate to the zero-range limit. In addition to being 
experimentally testable, our results impact the prospects of developing nuclear physics as a perturbation around the unitary limit. 
\end{abstract}


\maketitle

The study of strongly interacting, ultracold fermionic atoms has witnessed a large number of exciting developments, including the BEC-BCS crossover, the physics of polarons, optical lattices, three- or many-component experiments, and lower-dimensional systems, whether at zero or finite temperature~\cite{Braaten:2004rn,Giorgini:2008,Levinsen:2014,Naidon:2017}. Experimentally the systems are trapped fluids, where pairing correlations typically have a considerable impact on ground-state properties. Some of the most accurate theoretical studies of these systems have involved the use of Quantum Monte Carlo (QMC) or related methods. While early work focused on the two-component problem at or near the unitary regime~\cite{Carlson:2003,
Astrakharchik:2004,Stecher:2008,McNeil:2011,Bertaina:2011,Shi:2015,Anderson:2015,Galea:2016,Shi:2016,Lacroix:2016}, progress has also been made on the study of three- or more-fermionic components~\cite{Blume:2008,Dawkins:2017,Drut:2018}. A natural step is to tackle the four-component problem, where one could envision cold-atom experiments that directly probe the strongly interacting regime to address questions such as clustering in a many-particle system; crucially, a four-component cluster is bosonic in nature,
thereby increasing the likelihood that heavier systems will be bound.

Experiments with cold fermionic atoms hold the promise (in part already borne out~\cite{Horikoshi:2017}) of constraining aspects of nuclear physics which are not amenable to terrestrial experiments. Most obviously, pairing in two-component Fermi gases is directly related to the physics of neutron-star crusts~\cite{Gezerlis:2008,Gezerlis:2010,Carlson:2012,Gandolfi:2015,Lacroix:2017,Strinati:2018}. In low-density neutron matter, the components involved are neutrons with spin up ($n \uparrow$) and spin down ($n \downarrow$). Atomic nuclei involve two additional components of protons ($p \uparrow$, $p \downarrow$). Although the two-nucleon system is not exactly at unitarity, it has been argued that the properties of few-nucleon systems can be obtained in perturbation theory around this limit~\cite{Konig:2017,Konig:2016iny,Kolck:2017zzf,Kievsky:2018,Rupak:2018gnc,Gattobigio:2019,Konig:2019xxk}. Four-component fermionic systems thus might shed light on issues like clustering in light nuclei and the convergence to the thermodynamic limit, i.e., nucleonic matter.

In the unitary limit where the two-body scattering length diverges and the binding energy is vanishingly small, the two-body system is scale invariant and parameter-free. At or close to the unitarity limit, weakly bound systems can be described with an effective field theory (EFT), contact (or pionless) EFT~\cite{Hammer:2019poc}, where all interactions are of zero range. For bosons and three- or more-component fermions, the three-body system collapses ~\cite{Thomas:1935zz} unless a three-body interaction prevents it~\cite{Bedaque:1999a,Bedaque:1999b}; as a result, heavier systems will also collapse. The three-body force contains a single dimensionful parameter and gives rise to the Efimov effect~\cite{Efimov:1970zz} thanks to a remaining {\it discrete} scale invariance. Systems with more bosons also display the consequences of discrete scale symmetry~\cite{Platter:2004he,Hammer:2006ct,Schmidt:2009kq,Deltuva:2010xd,Stecher:2010,Nicholson:2012,Kievsky:2014,Yan:2015,Bazak:2016wxm,Bazak:2018qnu} and saturate at finite density~\cite{Carlson:2017}. 

While the four-nucleon system is well described in pionless EFT~\cite{Platter:2004zs,Stetcu:2006ey,Kirscher:2009aj,Lensky:2016djr,Konig:2017,Contessi:2017,Bansal:2018}, results for heavier nuclear systems ($^6$Li, $^{16}$O, $^{40}$Ca) have so far been somewhat disappointing: although binding energies are within the error of the leading-order calculation, systems are unstable with respect to breakup into $^{4}$He and $^{2}$H clusters~\cite{Stetcu:2006ey,Contessi:2017,Bansal:2018}. As one pushes pionless EFT to larger systems, the issue arises of whether the details of the interactions matter. In recent decades the strongest contender to account for these details has been chiral EFT~\cite{Hammer:2019poc}, where contact interactions are supplemented by exchanges of the lightest mesons (pions). Nuclear many-body calculations now routinely employ interactions from chiral EFT with much success (for example Refs.~\cite{Hebeler:2010,Gezerlis:2013,Coraggio:2013,Hagen:2014,Gezerlis:2014,Carbone:2014,Roggero:2014,Wlazlowski:2014,Soma:2014,
Elhatisari:2015,Tews:2016,Piarulli:2018,Lonardoni:2018,Lynn:2019,Stroberg:2019}), but they do not saturate properly at leading order~\cite{Ekstrom:2017koy,Sammarruca:2018bqh,Machleidt:2009bh}.
Properties of nuclear binding and clustering have recently been explored in the context of a lattice
approach using chiral or minimal interactions~\cite{Elhatisari:2016,Elhatisari:2017,Lu:2019}.

Here we consider  for the first time four-component unitary systems with more than four fermions. Like the corresponding bosonic systems~\cite{Carlson:2017}, these systems are expected to be universal in the sense that all energies are given by dimensionless numbers times the energy of the three-body ground state, $E_3$, or alternatively the four-body ground-state energy $E_4= 4.611 E_3$~\cite{Deltuva:2010xd}. We report on novel quantum Monte Carlo calculations where the antisymmetrization required to respect the Pauli exclusion principle is carried out explicitly. We use a variety of trial wave functions as we have had to explicitly check and minimize the effect of the fermion-sign problem on our results. We find that the form of the trial wave function plays a considerable role. Thankfully, our approach is variational, meaning that if a trial wave function leads to a low energy, we can discard other guesses for the trial wave function which gave higher values for the energy. 

Unlike bosonic systems, four is the maximum number of four-component fermions that can be found in relative $S$ waves. We might then expect clustering when an integer number of four particles are considered. We focus on the simplest such system, containing eight bodies. This is the analog of the $^8$Be nucleus, whose ground state is observed to be a narrow resonance very close to the two-$^4$He threshold~\cite{Benn:1967owz,Wustenbecker:1992}. As we describe in the remainder of this Letter, we find that eight four-component fermions at unitarity do cluster into two four-particle subsystems. This suggests that clustering is a universal feature of weakly bound, multicomponent fermion systems, with tantalizing implications to the wider program of producing nuclear observables as small corrections to the corresponding cold-atom ones. Since the binding energies of nuclei up to $^{52}$Fe are close to the energies of the corresponding number of independent alpha particles, a description of the eight-particle system holds the promise of being extensible to heavier systems.

Our Hamiltonian is:
\begin{equation}
\label{hamiltonian}
\hat{H} = -\frac{\hbar^2}{2m}\sum_i \nabla^2_i+\sum_{i<j}V_{i,j}+\sum_{i<j<k}V_{i,j,k}
\end{equation}
where $m$ is the particle mass, $V_{i,j}$ is a two-body attractive potential which acts between particles belonging to distinct components, and $V_{i,j,k}$ is a three-body repulsive potential where, again, each particle within a triplet must belong to a distinct component. At unitarity observables should be insensitive to the form of the potentials, as long as their ranges are small compared to interparticle distances. We take Gaussian forms with a common range $\mu^{-1}$:
\begin{equation}
\label{v2}
V_{i,j} = -V_2\mu^2\frac{2\hbar^2}{m} \exp [ -(\mu r_{ij})^2/2 ]
\end{equation}
\begin{equation}
\label{v3}
\begin{split}
V_{i,j,k} &= V_3 \Big( \frac{\mu}{4} \Big) ^2\frac{2\hbar^2}{m} \exp [ -(\mu R_{ijk}/4)^2/2 ]
\end{split}
\end{equation}
where  $r_{ij} = |\mathbf{r}_i - \mathbf{r}_j|$ and $R_{ijk} = (r_{ij}^2+r_{ik}^2+r_{jk}^2)^{-1/2}$.
The strengths $V_2$ and $V_3$ are adjusted to ensure, respectively, two-body unitarity and a nonzero four-body energy $E_4$; as a preliminary check, we ensured that our results for $E_4$ matched
the four-boson values from Ref.~\cite{Carlson:2017}. To produce dimensionless quantities, we employ $E_4$ or, alternatively, 
the corresponding length   $R_4=(-2mE_4/\hbar^2)^{-1/2}$.

To solve for the ground-state energy of this Hamiltonian we use a combination of the variational Monte Carlo (VMC) and diffusion Monte Carlo (DMC) methods. The VMC method is based on using trial wave functions that attempt to capture the physics of the system being studied. The VMC method evaluates the expectation value of the Hamiltonian using the trial wave functions by computing:
\begin{equation}
\label{vmcSUM}
E_V \approx \frac{1}{M}\sum^{M}_{i=1}E_L(\textbf{R}_i)
\end{equation}
where $E_L(\textbf{R}) = \psi^{-1}_T(\textbf{R}) \hat{H} \psi_T(\textbf{R})$ is the local energy,
$M$ is the number of sample points, and $\mathbf{R}$ (or $\mathbf{R}_i$)
encapsulates all the particle positions; this is the result of attacking a multidimensional integral
via the Monte Carlo approach. The trial wave functions $\psi_T$ contain variational parameters that are adjusted to find a lower VMC energy,
carried out via automated optimization techniques~\cite{Sorella:2001}. Once the VMC energy cannot be lowered further by adjusting these parameters we move onto the Diffusion Monte Carlo method. In DMC we are still making random walks in coordinate space
via so-called walkers. However, rather than evaluating a variational estimate of the ground-state energy, we are propagating through imaginary time to project to the ground state $|\psi_0\rangle$:
\begin{align}
\label{LargeTauLimit}
\ket{\Psi(\tau \rightarrow \infty)} &= \lim_{\tau \to \infty}e^{-(\hat{H}-E_T)\tau}\ket{\Psi(0)} \nonumber \\
&\propto \ket{\psi_0}\lim_{\tau \to \infty} e^{-(E_0-E_T)\tau}
\end{align}
where we have used imaginary time $\tau=it$ and have decomposed the trial wave function into the energy eigenstates using the completeness relation. The trial energy $E_T$ is included as an offset to the Hamiltonian. 
Of course, when propagating in imaginary time one has to also address the fermion-sign problem; we employ the fixed-node
approximation, which implies that our final DMC answers are upper bounds to the true ground-state energy.

In this work specifically, we study the ground-state energy of an eight-particle system. Such a system is made up of eight fermions, with two fermions each belonging to four components (\RomanNumeralCaps{1}, \RomanNumeralCaps{2}, \RomanNumeralCaps{3}, \RomanNumeralCaps{4}). 
We investigated three possibilities of increasing physical content, so we note ahead of time
that  Eq.~(\ref{TotPsi}) is the main new idea in this Letter.
 First, we extended the two-component BCS wave function (successfully used in DMC calculations of spin-$1/2$ Fermi 
gases~\cite{Carlson:2003}) to the problem of four components:

\begin{equation}
\label{BCS4spec}
\psi_T^A = f_J\big[\Phi^{ {\rm I}, {\rm II}}_{BCS}\Phi^{ {\rm III}, {\rm IV}}_{BCS}+\Phi^{ {\rm I}, {\rm III}}_{BCS}\Phi^{ {\rm II}, {\rm IV}}_{BCS}+\Phi^{ {\rm I}, {\rm IV}}_{BCS}\Phi^{ {\rm II}, {\rm III}}_{BCS}\big]
\end{equation}
where $\Phi^{m,n}_{BCS}$ is the BCS function that pairs components $m$ and $n$ and $f_J$ is a nodeless Jastrow function. $\Phi^{m,n}_{BCS}$ is simply the form used when a system contains only two components; Eq.~(\ref{BCS4spec}) applies it 
to all permutations of component pairing. The motivation behind this choice is the success of $\Phi^{m,n}_{BCS}$ in
describing the two-component system: via the use of 10 variational parameters, the two-component unitary Fermi gas is described extremely well~\cite{Carlson:2003,McNeil:2011}.

Second, we employed a ``cluster wave function,'' motivated by the forms used for $^8$Be in the mid-20th century~\cite{Wildermuth:1958}:
\begin{equation}
\label{FirstClusterWF}
\begin{split}
\psi_T^B = &\mathcal{A}\big[e^{-\beta_0 \sum_{\substack{i=1,3,5,7}}(\mathbf{r}_i-\mathbf{r}_{CM}^{1,3,5,7})^2} \times \\
&\qquad e^{-\beta_0 \sum_{\substack{j=2,4,6,8}}(\mathbf{r}_j-\mathbf{r}_{CM}^{2,4,6,8})^2} \times \\
&\qquad e^{-\beta_1(\mathbf{r}_{CM}^{1,3,5,7} - \mathbf{r}_{CM}^{2,4,6,8})^2}(\mathbf{r}_{CM}^{1,3,5,7} - \mathbf{r}_{CM}^{2,4,6,8})^n \big]
\end{split}
\end{equation} 
where $\beta_0$, $\beta_1$ and $n$ are variational parameters.
In this notation we have arbitrarily labeled the positions of the particles that belong to the component \RomanNumeralCaps{1} as $\mathbf{r}_1$ and $\mathbf{r}_2$, component \RomanNumeralCaps{2} has $\mathbf{r}_3$ and $\mathbf{r}_4$ and so on and we have written the center of mass for particles $a, b, c, d$ as $\mathbf{r}_{CM}^{a,b,c,d}$. In Eq.~(\ref{FirstClusterWF}), $\mathcal{A}$ is the antisymmetrization operator, needed to make the form of this wave function appropriately antisymmetric under exchange of identical particles, belonging to the same component. 
Here, clustering is captured in the first two exponential terms where a positive $\beta_0$ causes a decay in the wave function as particles move away from the center of mass of their cluster. Intercluster interactions are dictated by the remaining exponential and polynomial terms, and can be tuned with $\beta_1$ and $n$ to favor close mixing of the two clusters, or for the two to remain separate.

\begin{figure}[t]
\centering
\includegraphics[width=0.5\textwidth]{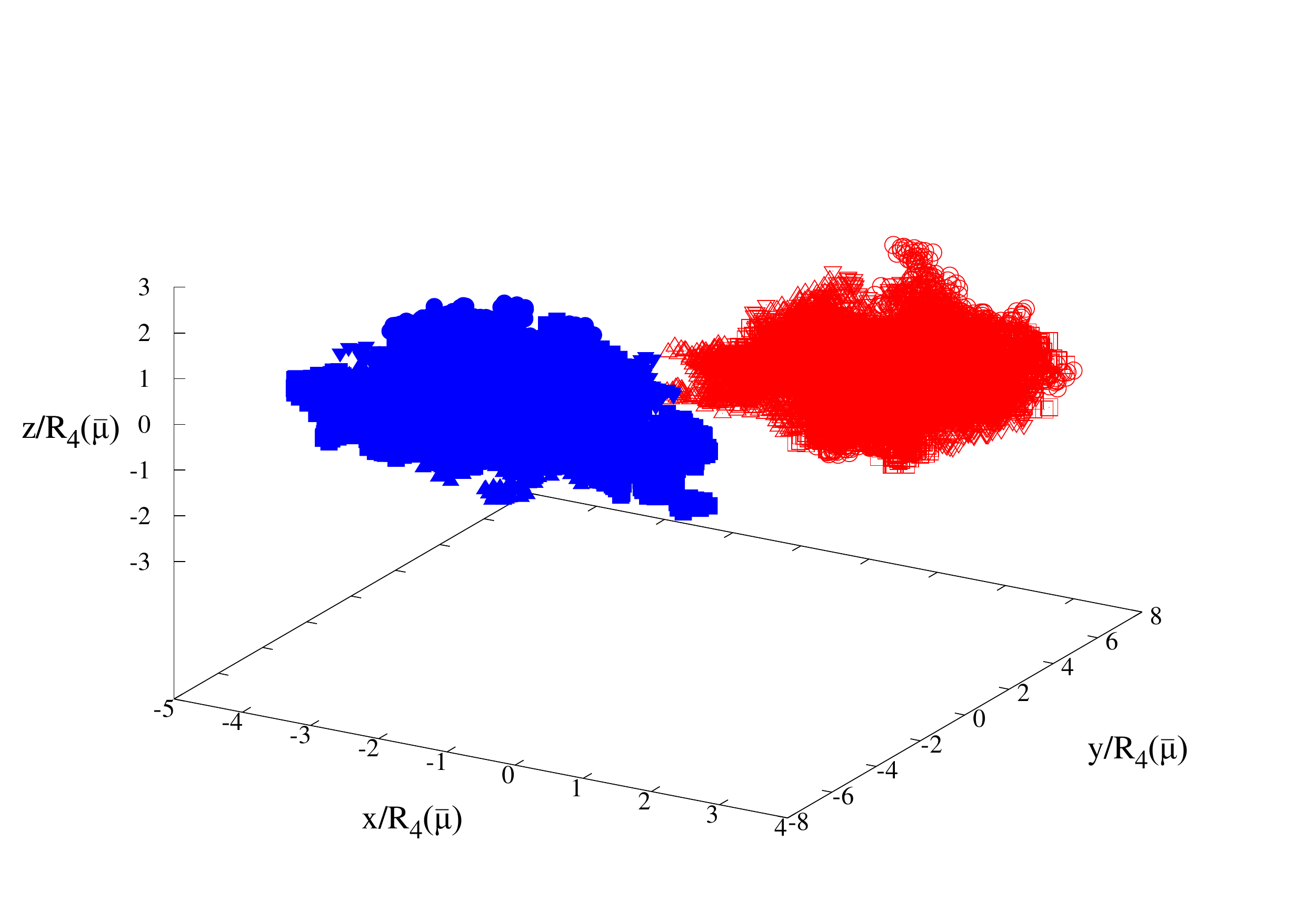} 
\caption{Particle positions over 5000 VMC steps, colored by how the particles have arranged themselves into clusters. Potential parameters are set to $V_3=3.0$ and $\bar\mu R_4(\bar\mu) =13.64$. The positions have been made dimensionless.}
\label{fig.Configurations}
\end{figure}

Third, we improved the trial wave function of Eq.~(\ref{FirstClusterWF}) by extending it to also 
allow for more complicated correlations: 
\begin{equation}
\label{TotPsi}
\begin{split}
\psi_T^C &= \mathcal{A}[F(\mathbf{r}_{CM}^{1,3,5,7}-\mathbf{r}_{CM}^{2,4,6,8}) \times f_J(\mathbf{r}_1,\mathbf{r}_3,\mathbf{r}_5,\mathbf{r}_7)\times \\
&\qquad f_J(\mathbf{r}_2,\mathbf{r}_4,\mathbf{r}_6,\mathbf{r}_8)\times \prod_{\substack{n=1,3,5,7 \\ m=2,4,6,8}} g(r_{nm})]
\end{split}
\end{equation}
where:
\begin{equation}
\label{F}
\begin{split}
F(\mathbf{r}_{CM}^{1,3,5,7}-\mathbf{r}_{CM}^{2,4,6,8})= 
\left (1-\gamma e^{- (\mathbf{r}_{CM}^{1,3,5,7}-\mathbf{r}_{CM}^{2,4,6,8})^2/\alpha^2} \right )^{-1}
\end{split}
\end{equation}
and:
\begin{equation}
\label{fc}
\begin{split}
f_J=\prod_i e^{-\alpha_{J} r_i^2}\prod_{i<j}\frac{K \tanh(\mu_Jr_{ij}) \cosh(\gamma_{J}r_{ij})}{r_{ij}}\\ \times \prod_{i<j<k}e^{u_Je^{-R_{ijk}^2/(2r_J^2)}}
\end{split}
\end{equation}
and, finally:
\begin{equation}
\label{g}
g(r_{nm})= \left (1-\gamma_g e^{-r_{nm}^2/\alpha_g^2} \right )^{-1}
\end{equation}
The 
$K$ and $\gamma_{J}$ are chosen such that the function $K \tanh(\mu_Jr_{ij}) \cosh(\gamma_{J}r_{ij})/r_{ij}$ goes to 1 and its derivative goes to 0 at $r_{ij}=d$, where $d$ is referred to as the ``healing distance.'' 
In this process, $\alpha$, $\gamma$, $\alpha_{J}$, $\mu_J$, $d$, $u_J$, $r_J$, $\gamma_g$ and $\alpha_g$ are variational parameters which are adjusted during VMC simulations to find an upper bound on the energy. 

\begin{figure}[t]
\centering
\includegraphics[width=0.45\textwidth]{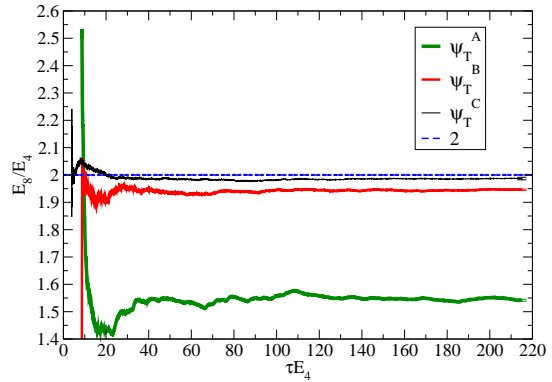} 
\caption{DMC running average for all three forms of trial wave functions. Potential parameters are set to $V_3=3.0$ and $\bar\mu R_4(\bar\mu) =13.64$.}
\label{fig.RunningAvs}
\end{figure}

The physical motivation behind $\psi_T^C$ was to capture the behavior seen in the four-particle boson case, which is equivalent to the four-particle fermion case when all fermions belong to distinct components. The $f_J$ functions are nodeless and have been used previously in simulations of bosonic clustering. In this context we are using this function to account for the formation of four-particle clusters (where each particle in the cluster belongs to each one of the distinct components), that we suspect will occur in the eight-particle case. Equations~(\ref{F}) and (\ref{g}) attempt to account for cross-cluster interactions, Eq.~(\ref{F}) is a function of the distance between the centers of mass of the two clusters, while Eq.~(\ref{g}) is simply a function of the separation distances between individual particles belonging to different clusters; both go to unity at large distances. 
The intercluster pair correlations allow for a deformation of the
individual clusters to enhance the attractive interaction between
different components between clusters, and to reduce the impact of
the Pauli repulsion between like particles. It is this additional 
correlation that might allow the system to bind. In a Born-Oppenheimer
picture, it also suggests that if two $N$-body clusters are bound (here
with $N=4$), problems with more than four components will also bind two
clusters since the ratio of attractive unlike interactions to Pauli
repulsion is increasing with the number of components.

We can qualitatively interpret the physical content of $\psi_T^C$ by plotting out the paths of the particles over some number of 
VMC steps. Figure~\ref{fig.Configurations} shows the system from the simulation using Eq.~(\ref{TotPsi}) over 5000 VMC steps,
taken after equilibration. We can see that the system forms two clusters, rather than a larger single cluster containing all of the particles; this is similar to what was seen in $^8$Be using the nuclear Green's function
Monte Carlo method~\cite{Wiringa:2000}. While the two
clusters here appear to be distinct, it is worth keeping in mind that the length scales involved in  $\psi_T^C$ are such
that the two regions communicate with each other; in the DMC method, the two clusters tend to drift apart. 
Note that in our case the two-body interactions have been tuned at unitarity, namely 
our calculations correspond to a deuteron with vanishing binding energy.
In the figure, we have divided all positions with 
$R_4(\bar\mu)$, the length corresponding to the range at which $\bar\mu R_4(\bar\mu) =13.64$.

\begin{figure}[t]
\centering
\includegraphics[width=0.45\textwidth]{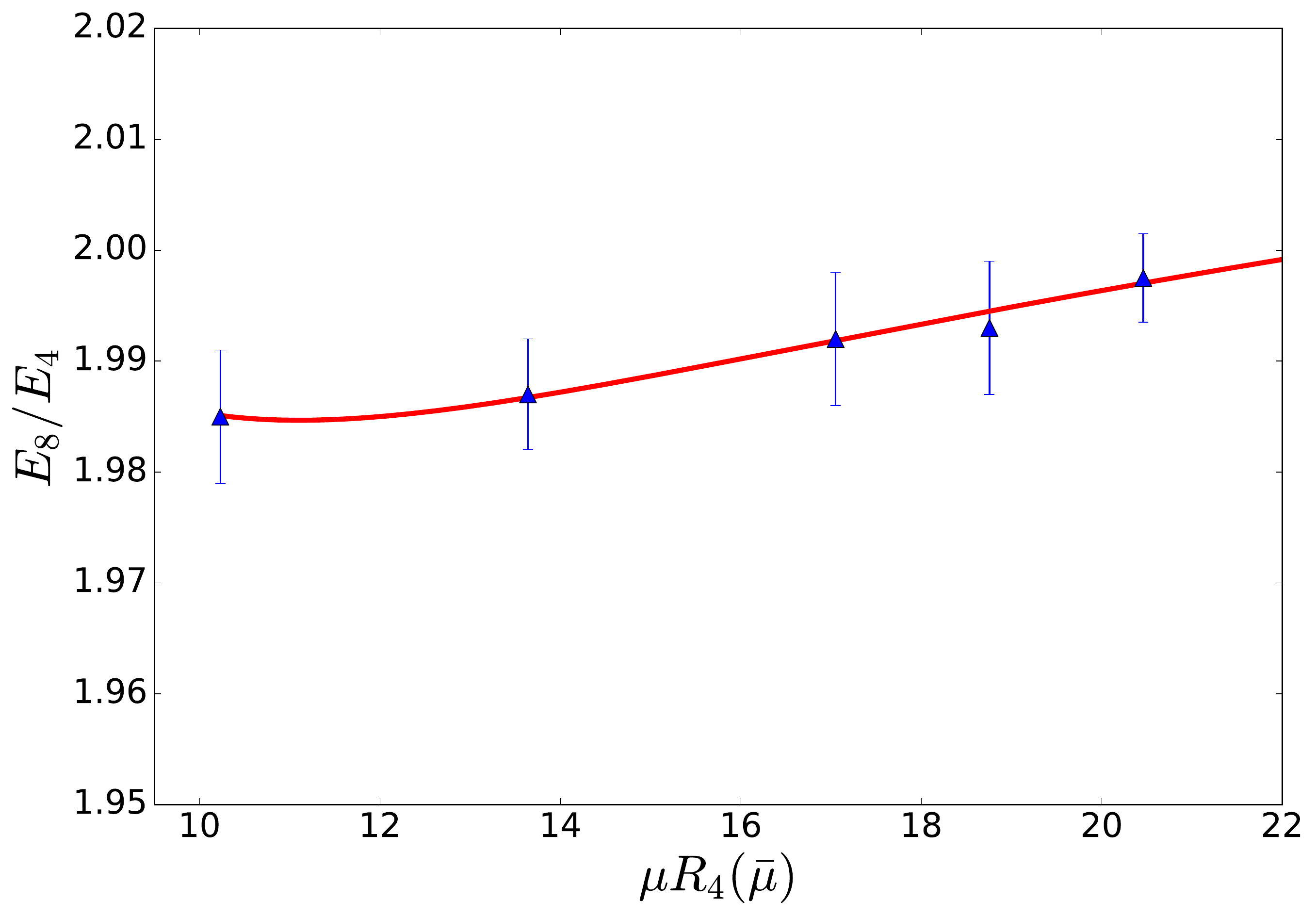} 
\caption{DMC results for the ratio of the eight-particle system energy to the four-particle system energy. The red curve is a fit 
meant to extrapolate to zero range.}
\label{fig.ratio}
\end{figure}

A simple benchmark for these calculations is to compare the energy calculated for the eight-particle system to that of a four-particle bosonic system with the same potential parameters. If our simulations have successfully reached the ground state of the eight-particle system 
we expect the energy found to be a factor of at least two times that of the energy of the corresponding four-particle system. We expect this upper limit because it should be possible for the system to produce two four-particle clusters (that are identical to those found in the four-particle simulation) which do not interact with one another, due to the tiny range of our potentials. If this is the case, the energy of the eight-particle system should be \textit{exactly} two times the four-particle system, due to the two independent energy contributions of each cluster. 

The results of this test are displayed in Fig.~\ref{fig.RunningAvs}. In this plot we see the running averages over imaginary time in DMC simulations for all three trial wave functions for a system where $V_3=3.0$ and $\bar\mu R_4(\bar\mu) =13.64$. We divide our $E_8$'s with the large-imaginary-time (central) value of $E_4$.
As can be seen in the figure, the running averages equilibrate over imaginary time and eventually converge to an answer with an acceptable statistical uncertainty. In this test, the trial wave function from Eq.~(\ref{TotPsi}) 
converged just below the $2E_4$ value. This test was carried out for multiple variations of adjusting the potential parameters and in all of these the same outcome was observed. More specifically, we varied the magnitude of $V_3$ (see the Supplemental Material~\cite{supple}): this impacts the value
of $E_4$, but each time we produced the same ratio $E_8/E_4$ within statistical error; this suggests that our findings are independent of the details of the short-range interaction.

It is worth highlighting that the computations employing $\psi_T^C$ were the only ones that led to an eight-particle system whose energy is one or two standard deviations away from breakup into two four-particle clusters. As a matter of fact, the runs with $\psi_T^A$ did not even come close to 
the $2E_4$ value: despite having ten variational parameters at its disposal, the BCS determinant is designed for a gas and therefore does not
do a good job of capturing clustering physics. The $\psi_T^B$ wave function, on the other hand, that \textit{is} cluster aware
does bring the energy to within 5\% of $2E_4$; of course, this is still noticeably different 
than the answer for two four-particle clusters, similarly to what was found in Ref.~\cite{Contessi:2017}. While physically we know that the system could dissolve into two disconnected
clusters, it is worth reiterating that DMC computations impose a nodal Ansatz: if this is sufficiently constricting, then 
the intuitively expected state of matter may not materialize. This is another way of saying that $\psi_T^C$ effectively
captures both the clustering physics and more involved correlations.

Once the testing of the trial wave function was completed, calculations were performed over a set of varying potentials. We are interested in the zero-range interaction limit, therefore in our simulations we proceeded to also 
vary $\mu$; we make this dimensionless by forming the product $\mu R_4(\bar\mu)$. In Fig.~\ref{fig.ratio} we plot the ratio of the energies of the eight-particle system and the four-particle system. We can see that as the range of the interactions become smaller, the absolute value of
the energy of the
eight-particle system becomes larger; this is consistent with the phase-shift expansion, which tells
us that the effective range tends to reduce the overall attraction.
 In order
to extrapolate, we fit our DMC results to the form:
\begin{equation}
\frac{E_8}{E_4} = c_0 + \frac{c_1}{\mu R_4(\bar\mu)} + \frac{c_2}{[\mu R_4(\bar\mu)]^2}~.
\end{equation}
In the limit of $\mu R_4(\bar\mu)$ going to infinity, namely a zero-range interaction, the ratio of the eight-particle to four-particle system goes to $2.04 \pm  0.05$ where we carried out standard error propagation. We also checked that employing a higher-degree (or lower-degree)
model does not qualitatively change the main result: $E_8$ is always within one standard deviation
of $2E_4$.

In summary, we have extended the DMC approach, which has in the past been applied to the two-component unitary
Fermi gas, to four-component unitary fermions. Employing microscopic interactions 
containing solely two- and three-body central potentials, we have investigated three different types of trial wave function, two of
which allow for clustering to emerge. For our most general form of the wave function, we have produced an eight-particle
state which is very close to decaying into four-particle clusters. These results constitute an example of 
pushing the applicability of pionless EFT to heavier systems. One could thus
add in a small perturbation which would bring $^8$Be to the physical point. More generally, our findings
could be experimentally tested in the future, when it becomes possible to manipulate four components of strongly
interacting fermionic atoms in the lab.

The authors wish to acknowledge several conversations with S. K\"onig and D. Lee.
This work was supported by the Natural Sciences and Engineering Research Council (NSERC) of Canada, the
Canada Foundation for Innovation (CFI), the Early Researcher Award (ERA) program of the Ontario Ministry of Research, Innovation and Science, the U.S. Department of Energy, Office of Science, Office of Nuclear Physics, under Contracts No. DE-AC52-06NA25396 and No. DE-FG02-04ER41338, and by the European Union Research and Innovation program Horizon 2020 under Grant No. 654002. 
Computational resources were provided by SHARCNET and NERSC.

\end{document}


\title{Supplemental Material for: Clustering of Four-Component Unitary Fermions}

\author{William G. Dawkins}
\affiliation{Department of Physics, University of Guelph, Guelph, Ontario N1G 2W1, Canada}
\author{J. Carlson}
\affiliation{Theoretical Division, Los Alamos National Laboratory, Los Alamos, New Mexico 87545, USA}
\author{U. van Kolck}
\affiliation{Institut de Physique Nucl\'eaire, CNRS-IN2P3, Universit\'e Paris-Sud, Universit\'e Paris-Saclay, 91406 Orsay, France}
\affiliation{Department of Physics, University of Arizona, Tucson, AZ 85721, USA}
\author{Alexandros Gezerlis}
\affiliation{Department of Physics, University of Guelph, Guelph, Ontario N1G 2W1, Canada}


\maketitle

The Hamiltonian used in this paper is:
\begin{equation}
\label{hamiltonian}
\hat{H} = -\frac{\hbar^2}{2m}\sum_i \nabla^2_i+\sum_{i<j}V_{i,j}+\sum_{i<j<k}V_{i,j,k}
\end{equation}
where
$V_{i,j}$ is a two-body attractive potential which acts between particles belonging to distinct components, and $V_{i,j,k}$ is a three-body repulsive potential where, again, each particle within a triplet must belong to a distinct component. 
Specifically, we take:
\begin{equation}
\label{v2}
V_{i,j} = -V_2\mu^2\frac{2\hbar^2}{m} \exp [ -(\mu r_{ij})^2/2 ]
\end{equation}
and:
\begin{equation}
\label{v3}
\begin{split}
V_{i,j,k} &= V_3 \Big( \frac{\mu}{4} \Big) ^2\frac{2\hbar^2}{m} \exp [ -(\mu R_{ijk}/4)^2/2 ]
\end{split}
\end{equation}
The strength $V_2$ is adjusted to ensure two-body unitarity.
In the main text, $V_3$ was kept fixed ($V_3=3.0$). 
The $\mu$ was initially fixed at $\bar\mu R_4(\bar\mu) =13.64$, but then varied for Fig.~3,
where we investigated the limit of $\mu R_4(\bar\mu)$ going to infinity, namely 
that of a zero-range interaction.

\begin{figure}[b]
\centering
\includegraphics[width=0.45\textwidth]{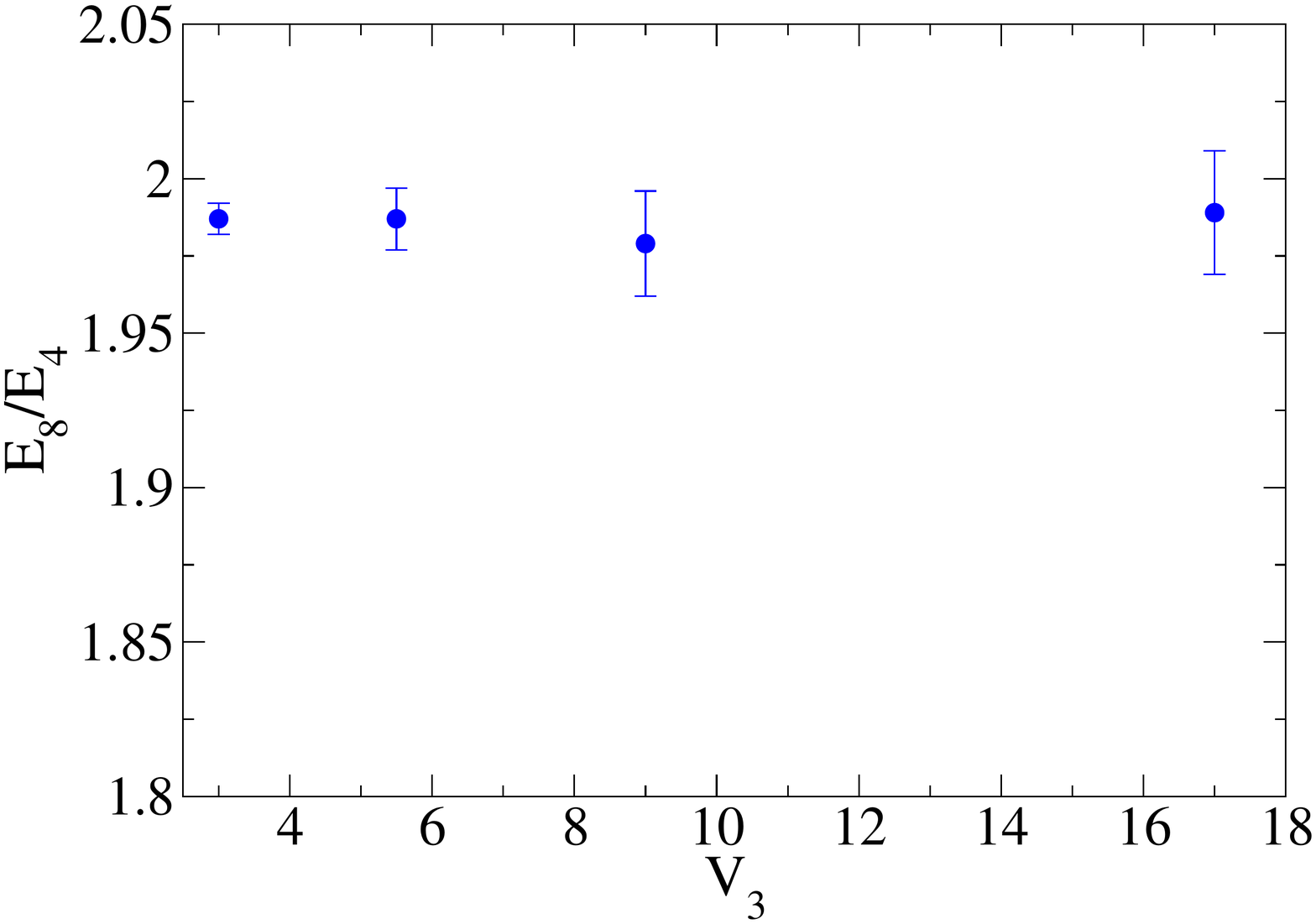} 
\caption{DMC results for the ratio of the eight-particle system energy to the four-particle system energy, 
as the strength of the three-body force is varied.}
\label{fig.withV3}
\end{figure}

Different choices of $V_3$
do impact the energy calculated for a given system. 
We have carried out Diffusion Monte Carlo (DMC)
calculations at several $V_3$ values; the results are shown in the figure on this page.
The larger values of $V_3$ lead to more repulsion and therefore a smaller four-particle binding energy.
Since the points on this plot are statistically indistinguishable, 
we see that, even though $E_4$ and $E_8$ depend on the specific value
$V_3$ has, $E_8/E_4$ does \textit{not} depend on $V_3$. 
This is further evidence that our findings are independent of the details of the short-range interaction.